# Longitudinal optical phonons in photonic time crystals containing a stationary charge

Sihao Zhang,[1,*] Junhua Dong,[1,*] Huanan Li,[1,†] Jingjun Xu,[1,‡] and Boris Shapiro[2,§]

[1]MOE Key Laboratory of Weak-Light Nonlinear Photonics, School of Physics, Nankai University, Tianjin 300071, China

[2]Department of Physics, Technion-Israel Institute of Technology, Haifa 32000, Israel

*Lorentzian-type media support optical phonons that oscillate with longitudinal polarization parallel to the wave direction, at a wave vector-independent frequency at which the permittivity becomes zero. Here, we study the interactions between the longitudinal optical phonons and Lorentzian medium-based dispersive photonic time crystals (PTCs). We demonstrate that a stationary charge embedded in the PTCs can excite these longitudinal modes through the conversion of the static polarization field induced by the charge. Furthermore, the PTCs can develop a momentum bandgap across the entire wave vector space to amplify the longitudinal modes. Remarkably, this infinite momentum bandgap can be established with minimal temporal modulation of the refractive index when creating the PTCs. Our approach expands the range of waves that can be manipulated in PTCs and shows potential for observing momentum bandgap phenomenon in realistic optical experiments, where the modulation depth of the refractive index is severely constrained.*

---

[*] These two authors contributed equally.
[†] hli01@nankai.edu.cn
[‡] jjxu@nankai.edu.cn
[§] boris@physics.technion.ac.il

Photonic time crystals (PTCs), a unique class of time-varying systems, represent an intriguing phase of matter that enables novel light-matter interactions [1]-[4]. Formed by applying periodic modulations to material parameters in a spatially uniform medium, PTCs preserve continuous space-translational symmetry while exhibiting discrete time-translational symmetry, resulting in a band structure characterized by distinct bands and bandgaps in momentum space ($\vec{k}$-space). Waves corresponding to a momentum bandgap can be exponentially amplified by drawing energy from the time-modulated system, leading to non-resonant gain effects without the need for phase-matching [3]. This phenomenon in PTCs has attracted considerable interest, both experimentally and theoretically [5]-[14]. Recent work [8]-[11] on behavior of sources (electric charges or dipoles), embedded into a PTC with momentum bandgaps, deserves special mention. It turns out that periodic temporal modulation of the medium can impel the embedded charge to radiate—a phenomenon with a potential for exotic tunable lasing.

In optics, creating a noticeable momentum bandgap demands ultrafast and significant changes—on the order of unity—in refractive index, posing challenges for many experimental platforms [15]-[16]. For instance, while conventional nonlinear optical materials enable nearly instantaneous index modulation, they typically achieve only a relative change of less than 1% [17]. Consequently, to date, PTCs and their associated momentum bandgap phenomenon have not yet been successfully implemented in optical regimes. To mitigate the need for large modulation depths in refractive index required to achieve a wide momentum bandgap, new approaches have recently emerged, including the use of resonant material systems [18] and the extension of wave characteristics through the incorporation of non-uniform waves [19].

In this paper, we explore longitudinal optical phonons in PTCs that utilize a Lorentz-type medium, modeled by a set of charged oscillators. The unique feature of such longitudinal modes is that their dispersion law does not contain the wave number $k$. We demonstrate that introducing a stationary charge allows for the longitudinal excitations at a time interface where the refractive index of the Lorentzian medium undergoes an abrupt change. We find that the PTCs, formed by a periodic arrangement of time interfaces, can sustain an infinite momentum bandgap for the longitudinal modes with only minimal relative modulation depth in the refractive index. Furthermore, this bandgap enables exponential amplification of longitudinal phonons in PTCs embedded with a stationary charge. Our results not only introduce a method for creating large momentum bandgaps with minimal refractive index modulation depth, but also significantly enrich time-varying photonics through their extreme interaction with longitudinal phonons.

***Longitudinal phonon oscillations and their excitation by a stationary charge at time interfaces*** — For simplicity, we focus on a lossless Lorentzian medium characterized by the (relative) permittivity $\varepsilon_r(\omega) = 1 + \omega_p^2/(\omega_0^2 - \omega^2)$, where $\omega_p$ and $\omega_0$ denote the plasma frequency and the intrinsic resonant frequency, respectively. A prominent example of such a medium is an ionic crystal which can support optical phonons coupled to the electromagnetic field. The zeros of $\varepsilon_r(\omega)$, occurring at $\omega = \pm\omega_L$ where $\omega_L = \sqrt{\omega_0^2 + \omega_p^2}$, define the oscillation frequencies for longitudinal optical phonons [20]. Notably, these longitudinal oscillations are independent of the wave vector $\vec{k}$ and do not couple with transverse electromagnetic waves within the bulk of the medium. In the very recent work of Feinberg et al. [21], this type of longitudinal mode has also been studied for plasmons within the context of plasmonic time crystals. To excite the longitudinal optical phonons, we introduce a stationary point charge $Q$ into the Lorentzian medium and examine its impact at a time interface associated with an abrupt change in the refractive index, $n(\omega) = \sqrt{\varepsilon_r(\omega)}$, or equivalently, the permittivity $\varepsilon_r(\omega)$. Specifically, we assume that the charge $Q$ is fixed at the origin $\vec{r} = 0$, and the permittivity $\varepsilon_r$ abruptly changes from $\varepsilon_r^{(1)}$ to $\varepsilon_r^{(2)}$ at time $t = 0$. Prior to the switch, the stationary charge generates a static electric field, resulting in a polarization of the medium described by $\vec{P}_S^{(1)} = \left(1 - \frac{1}{\varepsilon_r^{(1)}(0)}\right)\frac{jQ}{k^2}\vec{k}$ in the $\vec{k}$ space. It is well known that a time interface in an isotropic medium does not convert the static field from a stationary charge into radiation [22]. However, we demonstrate below that longitudinal phonon oscillations can still be generated in the medium when considering Lorentzian dispersion.

We note that after the abrupt switch of the refractive index, the polarization $\vec{P}_S^{(2)}$ due to the stationary charge becomes $\vec{P}_S^{(2)} = \left(1 - \frac{1}{\varepsilon_r^{(2)}(0)}\right)\frac{jQ}{k^2}\vec{k}$, while the associated magnetic flux density $\vec{B}_S = 0$ and the electric displacement $\vec{D}_S = \frac{jQ}{k^2}\vec{k}$ remain unchanged. Owing to preserved spatial uniformity, the wave vector $\vec{k}$ is conserved across the time interface [23], and the mismatch between the polarization vectors $\vec{P}_S^{(1)}$ and $\vec{P}_S^{(2)}$ for each $\vec{k}$ enables generation of longitudinal phonon oscillations. Indeed, at the time interface—triggered by an abrupt change in the resonant frequency $\omega_0$ from $\omega_0^{(1)}$ to $\omega_0^{(2)}$—the *total* polarization $\vec{P}$, its time derivative $\partial \vec{P}/\partial t$, and the *total* electric displacement $\vec{D}$ and magnetic flux density $\vec{B}$ all remain continuous, thereby defining temporal boundary conditions in the dispersive Lorentzian medium [24]-[25]. Furthermore, both the magnetic flux density $\vec{B}_L$ and the electric displacement $\vec{D}_L$ associated with the longitudinal optical phonons vanish, resulting in a polarization $\vec{P}_L$ generally expressed as:

$$\vec{P}_L(\vec{k}, t) = \{f(\vec{k}, t) + b(\vec{k}, t)\}\frac{jQ}{k^2}\vec{k}, \tag{1}$$

where $f(\vec{k},t) \propto e^{j\omega_L t}$ [and $b(\vec{k},t) \propto e^{-j\omega_L t}$] are the time-dependent complex amplitudes of the forward [and backward] waves for a given $\vec{k}$. As a result of the temporal boundary conditions, immediately after the switch, the polarization $\vec{P}_L(\vec{k},0^+)$ and its time derivative $\partial \vec{P}_L(\vec{k},0^+)/\partial t$ for the generated longitudinal phonons are given by $\vec{P}_L(\vec{k},0^+) = \vec{P}_S^{(1)} - \vec{P}_S^{(2)}$ and $\partial \vec{P}_L(\vec{k},0^+)/\partial t = 0$, respectively. These initial conditions, together with Eq. (1), allow us to determine the polarization $\vec{P}_L(\vec{k},t>0)$ for longitudinal phonons after switching time $t = 0$ as

$$\vec{P}_L(\vec{k},t>0) = \Delta \cos(\omega_L^{(2)} t) \frac{jQ}{k^2} \vec{k}, \qquad (2)$$

where $\Delta \equiv \left(\varepsilon_r^{(1)}(0) - \varepsilon_r^{(2)}(0)\right)/\left(\varepsilon_r^{(1)}(0)\varepsilon_r^{(2)}(0)\right)$, and the superscript $(m)$, with $m = 1, 2$, consistently denotes the states of the Lorentz medium characterized by the switched resonant frequencies $\omega_0^{(m)}$ and an unchanged plasma frequency $\omega_p$.

In Fig. 1(a), we examine the effect of a time interface [see the inset] on the longitudinal phonon generation in medium $m = 2$, utilizing Eq. (2). We fix $\omega_0^{(2)} \approx 0.89\omega_p$ so that the refractive index $n^{(2)}(0) = 1.5$ for medium $m = 2$, and vary $\omega_0^{(1)}$ of medium $m = 1$ to investigate the generated longitudinal mode $\vec{P}_L(\vec{k},0^+)$ immediately after switching. This mode is compared with the reference static component $\vec{P}_S^{(2)}(\vec{k})$, and is represented as $\vec{P}_L(\vec{k},0^+) = p_L(0^+)\vec{P}_S^{(2)}(\vec{k})$. As observed, the magnitude $|p_L(0^+)|$ vanishes in the absence of a time interface when $\omega_0^{(1)} = \omega_0^{(2)}$, and it increases but remains bounded (see dashed red lines) when $\omega_0^{(1)}$ deviates in either direction. By selecting $\omega_0^{(1)} \approx 0.58\omega_p$ [see brown star in Fig. 1(a)], such that the refractive index $n^{(1)}(0) = 2$ for medium $m = 1$, we demonstrate continuity across the time interface for both the total polarization $\vec{P}$ and its derivative $\partial \vec{P}/\partial t$ in Fig. 1(b). Before switching, the total polarization $\vec{P} = \vec{P}_S^{(1)}$, and after the switch, it transitions to a superposition of the excited longitudinal modes $\vec{P}_L(\vec{k}, t > 0)$ and the reference $\vec{P}_S^{(2)}$.

***Longitudinal phonon excitations by a stationary charge in PTCs***—The non-resonant gain effects of the momentum bandgap in PTCs, traditionally used to enhance the propagation of transverse electromagnetic waves, have recently been explored for potentially amplifying longitudinal plasmons, suggesting an unusually high optical gain [21]. Below, we examine how PTCs influence longitudinal phonon excitations in the presence of an embedded stationary charge. The PTCs we are considering are created by periodically switching the uniform Lorentz media between the states $m = 1$ and $m = 2$ after the initial time interface at $t = 0$, see Fig. **2**(a). We can fully describe the evolution of longitudinal phonons in the PTCs using the vector $\psi_L(\vec{k}, t) \equiv [f(\vec{k}, t), \quad b(\vec{k}, t)]^T$ (with the superscript $T$ denoting the transpose operation), which consists of the forward and backward time-dependent amplitudes $f(\vec{k}, t)$ and $b(\vec{k}, t)$ as specified in Eq. (1).

We determine the behavior of the vector $\psi_L(\vec{k}, t)$ over time by a generalized temporal transfer matrix formalism. Specifically, within a temporal slab of the medium at state $m = 1, 2$, longitudinal phonons experience free propagation, and thus the vector $\psi_L(\vec{k}, t)$ evolves according to the diagonal propagation matrix:

$$F^{(m)} = \text{diag}\{e^{j\omega_L^{(m)}T^{(m)}}, \quad e^{-j\omega_L^{(m)}T^{(m)}}\}, \qquad (3)$$

where $T^{(m)}$ represents the duration of the temporal slab. Correspondingly, the total polarization $\vec{P}(\vec{k}, t) = \vec{P}_L(\vec{k}, t) + \vec{P}_S^{(m)}(\vec{k})$ of the Lorentzian medium can be written as

$$\vec{P}(\vec{k}, t) = \left\{\left(1 - \frac{1}{\varepsilon_r^{(m)}(0)}\right) + [1, \quad 1]\psi_L(\vec{k}, t)\right\}\frac{jQ}{k^2}\vec{k}, \qquad (4)$$

which incorporates the contribution from the longitudinal phonons $\vec{P}_L(\vec{k},t)$ [see Eq. (1)] and those due to the stationary charge, namely $\vec{P}_S^{(m)}(\vec{k})$. The PTCs involve two types of time interfaces when the medium switches between states: from $m = 1$ to $m = 2$ and vice versa, from $m = 2$ to $m = 1$. In both cases, the temporal boundary conditions—specifically the continuity of the total polarization $\vec{P}$ and its time derivative $\partial \vec{P}/\partial t$—along with Eq. (4) for $\vec{P}(\vec{k},t)$ require that the vector $\psi_L(\vec{k},t)$ for longitudinal phonons undergoes an instantaneous variation. For the former transition, say at the switching time $t_S = T$ [see Fig. 2(a)], the rapid variation in the longitudinal phonons is given by $\psi_L(\vec{k},t_S^+) = J^{(2,1)}\psi_L(\vec{k},t_S^-) + S^{(2,1)}$, where the matching matrix $J^{(2,1)}$ and the source term $S^{(2,1)}$, essential for generating longitudinal phonons [see Eq. (2)], read

$$J^{(2,1)} = \frac{1}{2\omega_L^{(2)}}\begin{bmatrix} \omega_L^{(2)} + \omega_L^{(1)}, & \omega_L^{(2)} - \omega_L^{(1)} \\ \omega_L^{(2)} - \omega_L^{(1)}, & \omega_L^{(2)} + \omega_L^{(1)} \end{bmatrix}, \quad S^{(2,1)} = \frac{\Delta}{2}\begin{bmatrix} 1 \\ 1 \end{bmatrix}. \tag{5}$$

For the reverse transition, say at time $t_S = T^{(2)}$ [see Fig. 2(a)], $\psi_L(\vec{k},t_S^+) = J^{(1,2)}\psi_L(\vec{k},t_S^-) + S^{(1,2)}$ with $J^{(1,2)} = (J^{(2,1)})^{-1}$ and $S^{(1,2)} = -S^{(2,1)}$.

The propagation matrices $F^{(m)}$ for free propagation in medium $m = 1, 2$ [see Eq. (3)], the matching matrices $J^{(2,1)}$ and the source terms $S^{(2,1)}$ for the switching event from medium $m = 1$ to $m = 2$ [see Eq. (5)], along with their counterparts for reverse transition, constitute the building blocks of the generalized temporal transfer matrix method. This method enables us to determine the evolution of longitudinal phonon oscillations, represented by the vector $\psi_L(\vec{k},t)$, in PTCs containing a stationary charge. Essentially, each time interface excites a longitudinal phonon oscillation via its source term, and the excitations from different time interfaces evolve independently within the PTCs. Moreover, the evolution of each excitation is governed by the propagation matrices $F^{(m)}$ and the matching matrices $J^{(2,1)}$ and $J^{(1,2)}$, which dictate the dynamics following each event in the PTCs. As a result, after $n_t$ modulation cycles, the vector $\psi_L(\vec{k}, n_t T^+)$ for longitudinal phonons can be written as $\psi_L = \psi_L^{(2,1)} + \psi_L^{(1,2)}$, categorizing the excitations by type. Specifically, the first term $\psi_L^{(2,1)}$ originates from the excitation at all the time interfaces before time $t = n_t T^+$ where the medium transitions from state $m = 1$ to $m = 2$, and it is calculated as

$$\psi_L^{(2,1)} = (1 + M + \cdots + M^{n_t}) S^{(2,1)}, \tag{6}$$

where $M \equiv J^{(2,1)} F^{(1)} J^{(1,2)} F^{(2)}$ is the unit-cell temporal transfer matrix, describing the evolution of the excitation over one period [see dashed red box in Fig. **2**(a)]. Similarly, the second term $\psi_L^{(1,2)}$ relates to excitations at time interfaces when the medium switches from state $m = 2$ to $m = 1$, and it is given by

$$\psi_L^{(1,2)} = (1 + M + \cdots + M^{n_t - 1}) J^{(2,1)} F^{(1)} S^{(1,2)}. \tag{7}$$

By employing Eqs. (6) and (7), and noting that $S^{(1,2)} = -S^{(2,1)} = -\left(J^{(1,2)}\right)^{-1} S^{(2,1)}$, the vector $\psi_L(\vec{k}, n_t T^+)$ for longitudinal phonons after $n_t$ modulation cycles can be compactly expressed as

$$\psi_L(\vec{k}, n_t T^+) = \left\{ \frac{1 - M^{n_t + 1}}{1 - M} \left[1 - \left(F^{(2)}\right)^{-1}\right] + \left(F^{(2)}\right)^{-1} \right\} S^{(2,1)}. \tag{8}$$

***Infinite momentum bandgap of PTCs for longitudinal phonons*** — As seen from Eq. (8), the evolution of the generated longitudinal phonons in the PTCs with a stationary charge strongly depends on the unit-cell temporal transfer matrix $M$. Indeed, this matrix $M$ dictates the band structure for longitudinal phonons within the PTCs. Specifically, its eigenvalues $\lambda_\pm = \exp(j\Omega_\pm T)$ determine the Floquet frequencies $\Omega_\pm$, where the imaginary parts $\text{Im}(\Omega_\pm) = 0$ defines momentum bands for time-oscillating Floquet modes, and $\text{Im}(\Omega_\pm) \neq 0$ indicates bandgaps for modes that vary exponentially in time. The matrix $M$ here is independent of the wave vector $\vec{k}$, reflecting the $\vec{k}$-independent dispersion of longitudinal optical phonons. Consequently, the resulting momentum bandgap, once established, extends infinitely across the entire momentum space. This approach, potentially enabling an infinite momentum bandgap, somewhat resembles the method in Ref. [18] involving resonant materials, yet distinctively addresses longitudinal modes in a Lorentzian medium with a $\vec{k}$-independent longitudinal frequency $\omega_L$.

We proceed to investigate the development of the infinite momentum bandgap by examining the effects of the modulation frequency $\omega_M = 2\pi/T$ and the (relative) modulation depth $\delta \equiv (n^{(1)} - n^{(2)})/n^{(2)}$ for the refractive index $n \equiv n(0)$ on the Floquet frequencies $\Omega_\pm$. We note that the unit-cell temporal transfer matrix $M$ satisfies $\det M = 1$, and it can be parametrized as $M = \begin{bmatrix} a & d \\ d^* & a^* \end{bmatrix}$ with the asterisk denoting complex conjugate. As a result, the eigenvalues $\lambda_\pm$ can be expressed as $\lambda_\pm = \operatorname{Re} a \pm \sqrt{(\operatorname{Re} a)^2 - 1}$, and the momentum bandgap is induced when $(\operatorname{Re} a)^2 > 1$ for $|\lambda_+| = 1/|\lambda_-| \neq 1$, thereby ensuring $\operatorname{Im}(\Omega_+) = -\operatorname{Im}(\Omega_-) \neq 0$. For simplicity, we assume $T^{(1)} = T^{(2)} = T/2$ for the PTCs [see Fig. **2**(a)]. Utilizing the building blocks in Eqs. (3) and (5) for the matrix $M$, we can obtain the real part of its entry $a$, i.e., $\operatorname{Re} a$, as

$$\operatorname{Re} a = \cos\left(\frac{\pi \omega_L^{(1)}}{\omega_M}\right) \cos\left(\frac{\pi \omega_L^{(2)}}{\omega_M}\right) \\ - \frac{1}{2}\left(\frac{\omega_L^{(1)}}{\omega_L^{(2)}} + \frac{\omega_L^{(2)}}{\omega_L^{(1)}}\right) \sin\left(\frac{\pi \omega_L^{(1)}}{\omega_M}\right) \sin\left(\frac{\pi \omega_L^{(2)}}{\omega_M}\right), \quad (9)$$

where the longitudinal frequency $\omega_L^{(m)}$ for each medium $m = 1, 2$ is related to its refractive index $n^{(m)}$ by $\omega_L^{(m)}/\omega_p = n^{(m)}/\sqrt{(n^{(m)})^2 - 1}$. Except for differences in notation, Eq. (9) coincides with the result for longitudinal plasmons in Ref. [21]. As $\omega_L^{(1)}$ approaches $\omega_L^{(2)}$, corresponding to a vanishingly small modulation depth $\delta$ of the refractive index, the critical modulation frequences $\omega_M = 2\omega_L^{(2)}/N$, with $N$ as a positive integer, ensure $(\text{Re }a)^2 = 1$ according to Eq. (9) for the threshold of a momentum bandgap.

In Fig. **2**(b), we illustrate the strength of the momentum bandgap $\left|\text{Im}(\Omega_\pm)/\omega_p\right|$ in a density plot, as a function of the normalized modulation frequency $\omega_M/\omega_p$ and modulation depth $\delta$, near the critical frequency $\omega_M = 2\omega_L^{(2)}$ (see red star) where $\omega_L^{(2)} \approx 1.34\omega_p$ corresponds to the refractive index $n^{(2)} = 1.5$ for medium 2 as in Fig. **1**. As observed, the momentum bandgap $\text{Im}(\Omega_\pm) \neq 0$ develops over an extended range of the modulation frequency $\omega_M$ as the modulation depth $\delta$ increases, due to the major parametric resonance which imparts parametric gain to the collective longitudinal modes [26]. Furthermore, with a small detuning $\delta_M = \left(2\omega_L^{(2)} - \omega_M\right)/\omega_p > 0$ from the critical frequency $2\omega_L^{(2)}$ (see red star), there exists a threshold $\delta^{th}$ for the modulation depth $\delta$ that is required to achieve the infinite momentum bandgap. Up to the first order $O(\delta_M)$, the threshold value $\delta^{th}$ can be approximated as

$$\delta^{th} \approx \frac{\pi}{(2+\pi)} \frac{\left(n^{(2)}\right)^2 - 1}{\omega_L^{(2)}/\omega_p} \delta_M, \qquad (10)$$

based on $(\text{Re }a)^2 = 1$ (or specifically $\text{Re }a = -1$ for this case) and Eq. (9), see dashed red line in Fig. **2**(b). For each fixed detuning $\delta_M$, a leading-order perturbation of $\text{Im}(\Omega_\pm)/\omega_p$ around the threshold $\delta^{th}$ yields

$$\frac{\text{Im}(\Omega_\pm)}{\omega_p} \approx \pm \frac{1}{n^{(2)}} \sqrt{\frac{\delta_M}{\pi} \left(\frac{\omega_L^{(2)}}{\omega_p}\right)^{3/2}} \text{Re}\sqrt{\delta - \delta^{th}}, \qquad (11)$$

where the square root singularity indicates an exceptional point at which both the Floquet frequency $\Omega_\pm$ and their associated modes are degenerate. In Fig. 2(c), we plot $\text{Im}(\Omega_\pm)/\omega_p$ as a function of the modulation depth $\delta$ for various detunings $\delta_M = 0.001, 0.005$, and $0.01$. The dashed lines, based on the perturbation results from Eqs. (10) and (11), correspond closely with the numerical results (sold lines) near the threshold $\delta^{th}$. As the detuning $\delta_M$ decreases, bringing the modulation frequency $\omega_M$ closer to $2\omega_L^{(2)}$, the required modulation depth to produce the infinite momentum bandgap can be significantly reduced, albeit at the cost of a weaker bandgap strength. This behavior facilitates the observation of instabilities in the longitudinal modes of low-loss, time-modulated optical materials, where rapid but minimal modulation is feasible.

*Instabilities in longitudinal modes of PTCs with a stationary charge* — The optical gain exhibited in the momentum bandgap induces instability in the collective longitudinal modes of PTCs containing a stationary charge. Since the unit-cell temporal transfer matrix $M$ and the source term $S^{(2,1)}$ do not vary with the wave vector $\vec{k}$, $\psi_L(\vec{k}, n_t T^+)$ in Eq. (8) for generated longitudinal phonons in medium $m = 2$ remains constant across $\vec{k}$ space, i.e., $\psi_L(\vec{k}, n_t T^+) = \psi_L(n_t T^+)$. This uniformity simplifies the calculation of the total polarization $\vec{P}(\vec{r}, n_t T^+)$ in position space. By applying an inverse Fourier transform to $\vec{P}(\vec{k}, n_t T^+)$ from Eq. (4), we derive $\vec{P}(\vec{r}, n_t T^+)$ as the sum of the invariant static field $\vec{P}_S^{(2)}(\vec{r}) = \left(1 - \frac{1}{\varepsilon_r^{(2)}(0)}\right)\frac{Q}{4\pi r^3}\vec{r}$ in medium 2 and the generated longitudinal phonons $\vec{P}_L(\vec{r}, n_t T^+) = [1, \ 1]\psi_L(n_t T^+)\frac{Q}{4\pi r^3}\vec{r}$. In Fig. 3(a), using $\vec{P}_S^{(2)}(\vec{r})$ as a reference, we display the magnitude $|p_L(n_t T^+)|$ for generated longitudinal phonons $\vec{P}_L(\vec{r}, t) = p_L(t)\vec{P}_S^{(2)}(\vec{r})$ over the modulation cycles $n_t$, with a fixed modulation frequency detuning $\delta_M = 0.01$ and variable modulation depth $\delta$. The normalized value $p_L(t)$ here is equivalent to that of $\vec{P}_L(\vec{k}, t)$ relative to $\vec{P}_S^{(2)}(\vec{k})$ in $\vec{k}$ space, as introduced earlier [see Fig. **1**(a)]. As shown, the longitudinal phonons can be significantly amplified beyond a single time interface, with their growth rates as $n_t$ becomes large aligning with the bandgap strength [see dashed lines and also Fig. **2**(c)]. At a modulation depth of $\delta = 0.019$, Fig. **3**(b) depicts the time evolution of the longitudinal phonons $\vec{P}_L(\vec{r}, t)$ (green line), specifically $p_L(t)$, using the reference $\vec{P}_S^{(2)}(\vec{r})$. This data, obtained by numerically solving Maxwell's equations, matches well with theoretical predictions after each modulation cycle (circles) and is presented alongside the alternating static field $\vec{P}_S(\vec{r})$ (orange line) and the continuous total polarization $\vec{P}(\vec{r}, t)$ (blue line).

*Conclusions* — In this paper, we have explored the interactions between longitudinal optical phonons and a Lorentzian medium-based PTC, which is embedded with a stationary charge. Using the generalized temporal transfer matrix method, we have demonstrated that the dispersive PTC containing a stationary charge can not only excite but also amplify the longitudinal modes. We found that this amplification is governed by a momentum bandgap that extends across the entire momentum space. This infinite momentum bandgap originates from the unique, wave vector $(\vec{k})$-independent nature of longitudinal modes and can be established with minimal modulation depth of the refractive index, achievable in optical experimental platforms. Our approach enriches time-varying photonics by incorporating the exotic wave characteristics of longitudinal modes in dispersive PTCs. It holds potential for observing momentum bandgap phenomena using realistic, low-loss dielectrics and suggests new possibilities for controlling longitudinal phonons in bulk media. Finally, we emphasize that the stationary charge, embedded into our PTCs here, does not emit any radiation (i.e., no transverse electromagnetic waves) but only generates longitudinal excitations. The absence of radiation in an isotropic medium is in agreement with the general topological argument by Silveirinha [27].

*Acknowledgements* 一 This work was supported by the National Natural Science Foundation of China (Grant Nos. 12274240, 92250302).

**Figures**

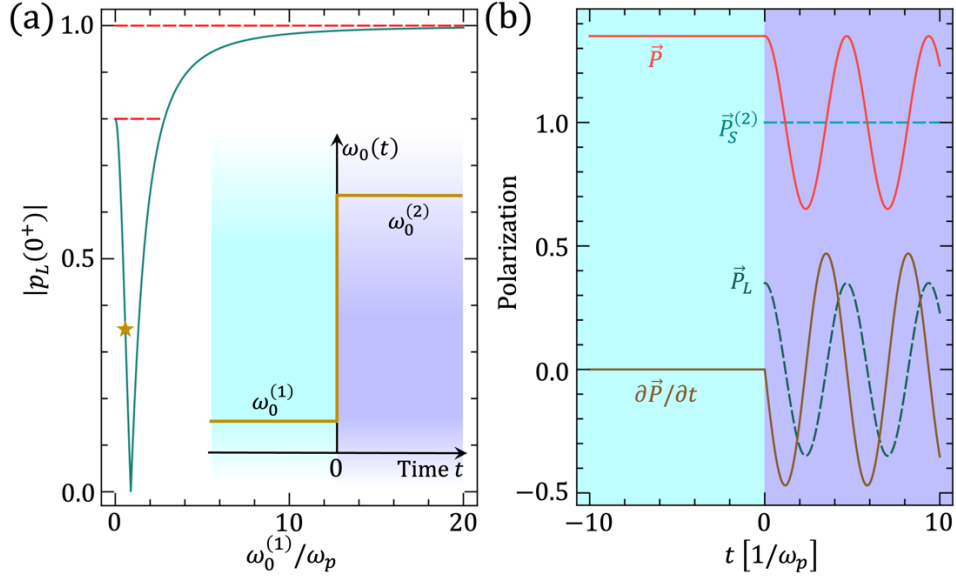

**Fig. 1.** (a) Magnitude of $p_L(0^+)$ for the generated longitudinal mode in medium 2, plotted against the resonant frequency $\omega_0^{(1)}$ of the medium 1 before the time interface (see inset). The resonant frequency $\omega_0^{(2)}$ for medium 2 after the time interface is fixed at $\omega_0^{(2)} \approx 0.89\omega_p$, setting its refractive index at $n^{(2)}(0) = 1.5$. The limiting values of $|p_L(0^+)|$ as $\omega_0^{(1)}$ approaches 0 and $\infty$ are marked by dashed red lines. (b) Time evolution of the normalized total polarization $\vec{P}$ and its derivative $\partial \vec{P}/\partial t$, along with the generated longitudinal mode $\vec{P}_L$ and the reference $\vec{P}_S^{(2)}$ after the time interface, corresponding to $\omega_0^{(1)} \approx 0.58\omega_p$ for medium 1 with a refractive index $n^{(1)}(0) = 2$ [see brown star in (a)].

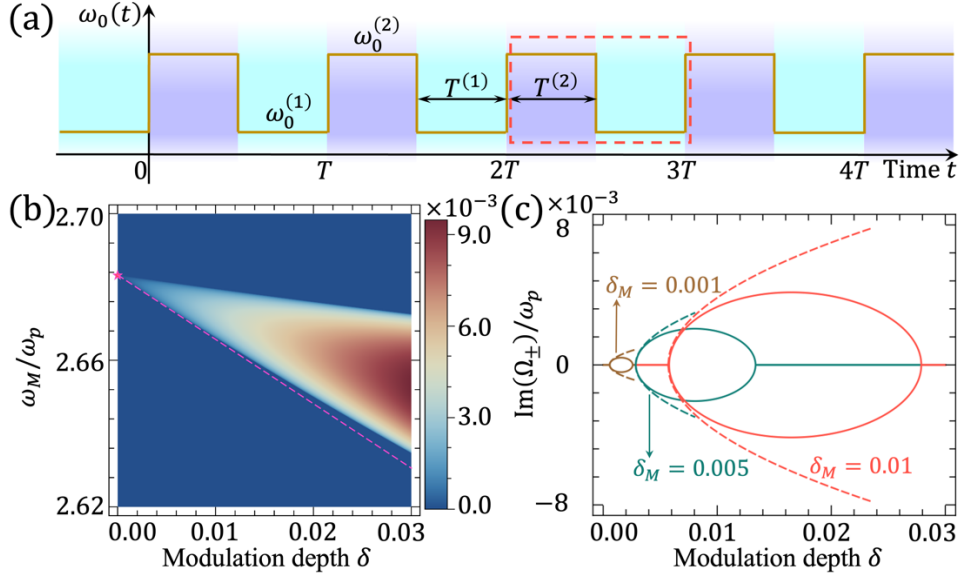

**Fig. 2.** (a) Schematic of the PTCs considered in the text. (b) Density plot for the strength $|\text{Im}(\Omega_\pm)/\omega_p|$ of the infinite momentum bandgap as a function of the normalized modulation frequency $\omega_M/\omega_p$ and modulation depth $\delta$. The red star marks the critical frequency $2\omega_L^{(2)}$ as $\delta$ approaches zero, while the dashed red line represents the predicted modulation depth threshold $\delta^{th}$ for the momentum bandgap, see Eq. (10). (c) $\text{Im}(\Omega_\pm)/\omega_p$ plotted against modulation depth $\delta$ for various detunings $\delta_M = 0.001, 0.005$, and $0.01$ from the critical frequency $2\omega_L^{(2)}$ [indicated by red star in (b)]. Solid lines denote numerical results, while dashed lines represent calculations from Eq. (11) using perturbation methods. Parameters for both (b) and (c) include $T^{(1)} = T^{(2)} = T/2$, and a fixed refractive index of $n^{(2)} = 1.5$ for medium 2, resulting in $\omega_0^{(2)} \approx 0.89\omega_p$ and $\omega_L^{(2)} \approx 1.34\omega_p$ for the PTCs in (a).

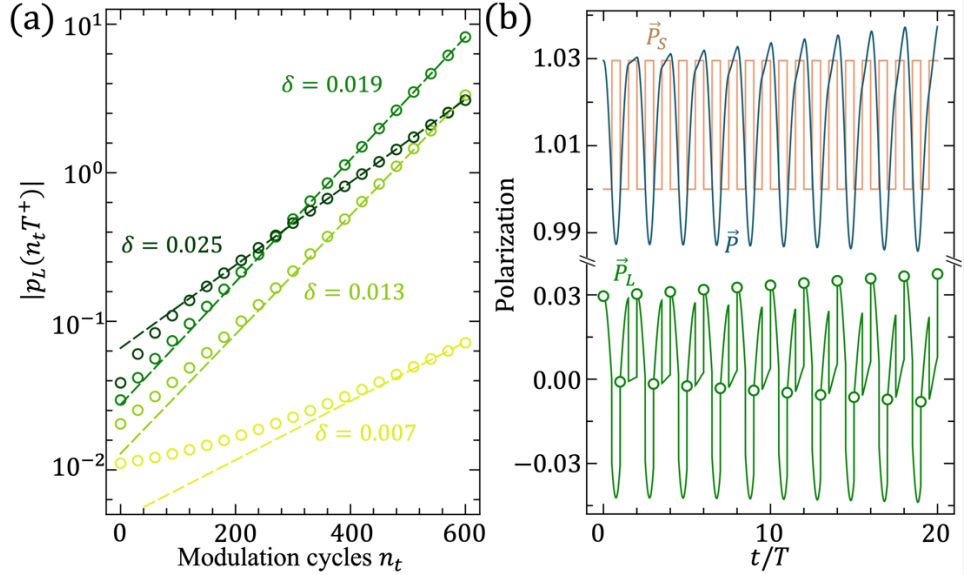

**Fig. 3.** (a) Magnitude of $p_L(n_t T^+)$ for the generated longitudinal phonons in medium 2, plotted against the modulation cycles $n_t$ for various modulation depth $\delta = 0.007, 0.013, 0.019$ and $0.025$. The detuning from the critical frequency $2\omega_L^{(2)}$ is fixed at $\delta_M = 0.01$ [see Fig. **2**(c)], and the growth rates derives from the bandgap strength are depicted by dashed lines for each $\delta$. (b) Time evolution of the normalized total polarization $\vec{P}$, longitudinal mode $\vec{P}_L$ and the alternating static field $\vec{P}_S$, all relative to the reference $\vec{P}_S^{(2)}$, for the case of $\delta = 0.019$ in (a). Empty circles represent theoretical predictions of $\vec{P}_L$ at time $t = n_t T^+$. Other parameters for both (a) and (b) are the same as in Fig. **2**(b, c).